\documentclass[oupdraft]{bio}

\usepackage{amsmath,amssymb}
\usepackage{booktabs}
\usepackage{rotating}
\usepackage{multirow}
\usepackage{graphicx}
\usepackage{color}
\usepackage{subcaption}
\usepackage{cite}

\DeclareMathOperator*{\argmaxC}{\arg\max}

\def\bSig\mathbf{\Sigma}

\numberwithin{equation}{section}
\theoremstyle{plain}

\begin{document}

\title{Copula-based Semiparametric Regression Method for Bivariate Data under General Interval Censoring}

\author
{TAO SUN, YING DING$^\ast$ \\[4pt]
\textit{Department of Biostatistics,
University of Pittsburgh,
Pittsburgh, PA,
U.S.A.}
\\[2pt]
{yingding@pitt.edu}}

\markboth%
{T. SUN and Y. DING}
{Copula semiparametric model for bivariate interval-censored data}

\maketitle

%\footnotetext{To whom correspondence should be addressed.}

\begin{abstract}
{This research is motivated by discovering and underpinning genetic causes for the progression of a bilateral eye disease, Age-related Macular Degeneration (AMD), of which the primary outcomes, progression times to late-AMD, are bivariate and interval-censored due to intermittent assessment times. We propose a novel class of copula-based semiparametric transformation models for bivariate data under general interval censoring, which includes the case 1 interval censoring (current status data) and case 2 interval censoring. Specifically, the joint likelihood is modeled through a two-parameter Archimedean copula, which can flexibly characterize the dependence between the two margins in both tails. The marginal distributions are modeled through semiparametric transformation models using sieves, with the proportional hazards or odds model being a special case. We develop a computationally efficient sieve maximum likelihood estimation procedure for the unknown parameters, together with a generalized score test for the regression parameter(s). For the proposed sieve estimators of finite-dimensional parameters, we establish their asymptotic normality and efficiency. Extensive simulations are conducted to evaluate the performance of the proposed method in finite samples. Finally, we apply our method to a genome-wide analysis of AMD progression using the Age-Related Eye Disease Study (AREDS) data, to successfully identify novel risk variants associated with the disease progression. We also produce predicted joint and conditional progression-free probabilities, for patients with different genetic characteristics.}
{Bivariate; Copula; GWAS; Interval-censored; Semiparametric; Sieve.}
\end{abstract}

\section{Introduction} \label{sec:intro}
Bivariate time-to-event endpoints are frequently used as co-primary outcomes in biomedical and epidemiological fields. For example, two time-to-event endpoints are often seen in clinical trials studying the progression (or recurrence) of bilateral diseases (e.g., eye diseases) or complex diseases (e.g., cancer and psychiatric disorders). The two endpoints are correlated as they come from the same individual. Bivariate interval-censored data arise when both events are not precisely observed due to intermittent assessment times. Therefore, the event times are only known to belong to an interval (i.e., case 2 interval-censored). A further complication is that the event status can be indeterminate (i.e., right-censored) for individuals who are event-free at their last assessment time. The special case when there exists only one assessment time, leading to the bivariate current status data (events are either left- or right-censored), can also happen for some individuals. Therefore, the bivariate data we are interested in modeling are under general interval censoring, which may include a mixture of left-, right- and interval-censored data.

Our motivating example of such bivariate general interval-censored data came from a large clinical trial studying the progression of a bilateral eye disease, Age-related Macular Degeneration (AMD), of which the two-eyes from the same patient were periodically examined for late-AMD. The study aims to discover genetic variants that are significantly associated with AMD progression, as well as to characterize both the joint and conditional risks of AMD progression. For example, the joint 5-year progression-free probability for both eyes is a clinically significant measure to group patients into different risk categories. Similarly, for patients who have one eye already progressed, the conditional 5-year progression-free probability for the non-progressed eye (given its fellow eye already progressed) is vital to both clinicians and patients. Therefore, a desired statistical method needs to characterize and predict both joint and conditional risk profiles and assess the covariate effects on them.

There are several approaches to model bivariate interval-censored data. For example, \citet{bivariate_cox_marginal}, \citet{kim2002analysis}, \citet{chen2007proportional}, \citet{tong2008regression} and \citet{chen2013linear} fitted various marginal models for multivariate interval-censored data. All these approaches model the marginal distributions based on the working independence assumption, and thus cannot produce joint or conditional distributions. Another popular method is based on frailty models (for example, \citeauthor{frailty_oaks}, \citeyear{frailty_oaks}), which are mixed effects models with a latent frailty variable applied to the conditional hazard functions. For example, \citet{chen2009frailty} and \citet{chen2014analysis} built frailty proportional hazards (PH) models with piecewise constant baseline hazards for multivariate current status data and interval-censored data, respectively. \citet{wen_chen2013} and \citet{wang2015regression} developed Gamma-frailty PH models for bivariate interval-censored data through a nonparametric maximum likelihood estimation approach and bivariate current status data through a sieve estimation approach, respectively. Recently, \citet{frailty_case_II_transformation_sieve} and \citet{frailty_case_II_transformation_NPMLE} proposed frailty-based transformation models for bivariate or multivariate interval-censored data, and obtained parameter estimates through the sieve maximum likelihood estimation and nonparametric maximum likelihood estimation, respectively. For frailty models, the covariate effects are typically interpreted on the conditional level by conditioning on the random frailty term.

The third popular approach is based on copula models \citep[for example]{Clayton}. Unlike the marginal or frailty approaches, the copula-based methods directly connect the two marginal distributions through a copula function to construct the joint distribution, of which the copula parameter determines the dependence. This unique feature makes the modeling of the margins separable from the copula function, which is attractive from both the modeling perspective and the interpretation purpose. Both joint and conditional distributions can be obtained from copula models. Several copula models have been proposed in the literature. \citet{copula_case_I_ph} used sieve estimation in a copula model with PH margins for bivariate current status data. \citet{copula_case_I_ph_piecewise} and \citet{copula_case_II_ph_pw} developed estimating equations for copula models with piecewise constant baseline marginal hazards for clustered current status and interval-censored data, respectively. \citet{HuCaseICopulaPH_2017} developed a semiparametric sieve approach for bivariate current status data using copula framework with PH margins. To date, most copula-based regression models only handle a specific interval censoring type and are often limited to the PH assumption. Also, the most frequently used copula models, such as Clayton, Gumbel, and Frank, all use only one dependence parameter, which can be lack of flexibility.

\citet{copula_vs_frailty} and \citet{wienke2010frailty} have discussed the connection and distinction between copula and frailty models. For example, the Clayton copula has the same mathematical expression as the Gamma frailty model in terms of the joint survival distribution. However, their marginal survival functions are modeled differently. Specifically, the marginal function under the Clayton model only involves the time and covariate effects, whereas the marginal function under the Gamma frailty model includes time and covariate effects, and also the frailty parameter. As a result, the joint distribution functions of the Clayton copula and Gamma frailty models are not equivalent, except when the two margins are independent. More details are discussed in the Appendix C of Supplementary Materials. In this paper, the objectives of our real study lead us to choose copula-based models, which offer a straightforward interpretation of covariate effects and dependence strength, as well as an easy generation of joint and conditional survival distributions.

We propose a class of copula-based semiparametric transformation model for bivariate data subject to general interval censoring. For the copula model, we use a two-parameter copula function that can flexibly handle dependence structure on both upper and lower tails, and the dependence strength can be quantified via Kendall's $\tau$. For the marginal model, we use the semiparametric transformation model that incorporates a variety of models including PH and PO models. We approximate the infinite-dimensional nuisance parameters using sieves with Bernstein polynomials and propose a novel maximum likelihood estimation procedure which is computationally stable and efficient. We establish the asymptotic normality and efficiency for the sieve estimators of finite-dimensional model parameters. Moreover, we develop a computationally efficient generalized score test with numerical approximations of the score function and observed Fisher information for testing a large number of covariates (e.g., millions of SNPs). Lastly, the joint distribution can be directly obtained from our model, making it applicable to estimating the joint and conditional progression profiles for patients with different characteristics.

The paper is organized as follows. Section \ref{sec_notation_likelihood} introduces the model and the joint likelihood function. Section \ref{sec_estimation_test_procedure} presents the sieve maximum likelihood estimation procedure, the asymptotic properties, and the generalized score test. Section \ref{sec_simulations} illustrates extensive simulation studies for the estimation and testing performances of our proposed methods. We analyze the AREDS data and present the findings in Section \ref{sec_real_data}. Finally, we discuss and conclude in Section \ref{sec_conclusions}. Additional simulation and analysis results, the regularity conditions, proofs and additional technical details are provided in the Supplementary Materials.

\section{Notation and Likelihood} \label{sec_notation_likelihood}
\subsection{Copula model for bivariate censored data} \label{subsec_copula}
Assume there are $n$ independent subjects in a study. For subject $i$, we observe $D_i=\{(L_{ij},R_{ij},Z_{ij}), \\ j = 1,2\}$, where $(L_{ij}, R_{ij}]$ is the time interval that the true event time $T_{ij}$ lies in and $Z_{ij}$ is the covariate vector. When $R_{ij} = \infty$, $T_{ij}$ is right-censored, and when $L_{ij} = 0$, $T_{ij}$ is left-censored. We define the marginal survival function for subject $i$ margin $j$ as $S_{j}(t_{ij} | Z_{ij})=pr(T_{ij} > t_{ij} | Z_{ij})$ and the joint survival function for subject $i$ as $S(t_{i1},t_{i2} | Z_{i1}, Z_{i2})=pr(T_{i1} > t_{i1}, T_{i2} > t_{i2} | Z_{i1}, Z_{i2})$.

By the Sklar's theorem (\citeauthor{sklar}, \citeyear{sklar}), so long as marginal survival functions $S_j$ are continuous, there exists a unique function $C_\eta$ that connects two marginal survival functions into the joint survival function: $S(t_1,t_2 | Z_1, Z_2) = C_\eta(S_1(t_1 | Z_1),S_2(t_2 | Z_2)),\ t_1, t_2 \geq 0.$ Here, the function $C_\eta$ is called a copula, which maps $[0,1]^2$ onto $[0,1]$ and its parameter $\eta$ measures the dependence between the two margins. A signature feature of the copula is that it allows the dependence to be modeled separately from the marginal distributions \citep{Nelson_2006}.

One favorite copula family for bivariate censored data is the Archimedean copula family, which usually has an explicit formula. Two frequently used Archimedean copulas are the Clayton (\citeauthor{Clayton}, \citeyear{Clayton}) and Gumbel (\citeauthor{gumbel}, \citeyear{gumbel}) copula models, which account for the lower or upper tail dependence between two margins using a single parameter.

Here, we consider a more flexible two-parameter Archimedean copula model \citep{Joe}, which is formulated as
\begin{eqnarray}
C_{\alpha,\kappa}(u,v)=[1+\{(u^{-1/\kappa}-1)^{1/\alpha} + (v^{-1/\kappa}-1)^{1/\alpha} \}^{\alpha}]^{-\kappa}, \ \alpha \in (0,1], \ \kappa \in (0,\infty), \label{two-para-c}
\end{eqnarray}
where $u$ and $v$ are two uniformly distributed margins. The two dependence parameters ($\alpha$ and $\kappa$) account for the correlation between $u$ and $v$ at both upper and lower tails, and they explicitly connect to the Kendall's $\tau$ with $\tau = 1- {2\alpha\kappa}/(2\kappa + 1)$. In particular, when $\alpha = 1$, the two-parameter copula (\ref{two-para-c}) becomes the Clayton copula, and when $\kappa \rightarrow \infty$, it becomes the Gumbel copula. Thus, the two-parameter copula model provides more flexibility in characterizing the dependence than the Clayton or Gumbel copula.

\subsection{Joint likelihood for bivariate data under general interval censoring} \label{subsec_joint_lik}
Based on the notation introduced in Section \ref{subsec_copula}, the joint likelihood function using the two-parameter copula model can be written as
\begin{eqnarray}
\label{joint_likelihood}
&&L_{n}(S_1, S_2, \alpha,\kappa \mid D) = \prod_{i=1}^n pr(L_{i1} < T_{i1} \leq R_{i1}, L_{i2} < T_{i2} \leq R_{i2} \mid Z_{i1}, Z_{i2}) \nonumber \\
& = & \prod_{i=1}^n \biggl\{pr(T_{i1} > L_{i1}, T_{i2} > L_{i2} \mid Z_{i1}, Z_{i2}) - pr(T_{i1} > L_{i1}, T_{i2} > R_{i2} \mid Z_{i1}, Z_{i2}) \nonumber \\
&& \quad - pr(T_{i1} > R_{i1}, T_{i2} > L_{i2} \mid Z_{i1}, Z_{i2}) + pr(T_{i1} > R_{i1}, T_{i2} > R_{i2} \mid Z_{i1}, Z_{i2}) \biggl\}  \nonumber \\
& = & \prod_{i=1}^n\biggl[C_{\alpha,\kappa}\{S_1(L_{i1}\mid Z_{i1}),S_2(L_{i2}\mid Z_{i2})\} - C_{\alpha,\kappa}\{S_1(L_{i1}\mid Z_{i1}),S_2(R_{i2}\mid Z_{i2})\} \nonumber \\
&& \quad - C_{\alpha,\kappa}\{S_1(R_{i1}\mid Z_{i1}),S_2(L_{i2}\mid Z_{i2})\} + C_{\alpha,\kappa}\{S_1(R_{i1}\mid Z_{i1}),S_2(R_{i2}\mid Z_{i2})\} \biggr].
\end{eqnarray}
For a given subject $i$, if $T_{ij}$ is right-censored, then any term involving $R_{ij}$ becomes 0 (since $R_{ij}$ is set to be $\infty$). Then the joint survival function for subject $i$ reduces to either only one term (if both $T_{i1}$ and $T_{i2}$ are right-censored) or two terms (if one $T_{ij}$ is right-censored). The particular case of current status data can also fit into this model frame, where either $L_{ij}$ is 0 (if the event has already occurred before the examination time, which is $R_{ij}$ in this case) or $R_{ij}$ is $\infty$ (if the event has not happened upon the examination time, which is $L_{ij}$ in this case). Therefore, the likelihood function (\ref{joint_likelihood}) can handle the general form of bivariate interval-censored data.

\subsection{Semiparametric linear transformation model for marginal survival functions} \label{subsec_semipar}
We consider the semiparametric transformation models for marginal survival functions:
\begin{equation}
S_{j}(t\mid Z_j) = \exp[-G_j\{\exp(Z_j^{T}\beta_j)\Lambda_{j}(t)\}], \ j = 1,2, \label{tran_mod}
\end{equation}
where $G_j(\cdot)$ is a pre-specified strictly increasing function, $\beta_j$ is a vector of unknown regression coefficients, and $\Lambda_{j}(\cdot)$ is an unknown non-decreasing function of $t$. In model (\ref{tran_mod}), the transformation function $G_j(\cdot)$, the regression parameter $\beta_j$ and the infinite-dimensional parameter $\Lambda_j(\cdot)$ are all denoted as margin-specific (indexed by $j$) for generality. In practice, some or all of them can be the same for the two margins, and in that case, the corresponding index $j$ can be dropped.

This model (\ref{tran_mod}) contains a class of survival models. For example, when $G(x) = x$, the marginal survival function follows a PH model. When $G(x) = \log(1+x)$, the marginal function becomes a proportional odds (PO) model. In practice, the transformation function can also be ``estimated'' by the data. For example, the commonly used Box-Cox transformation $G(x) = \{(1+x)^r -1\}/r$, $r > 0$, or the logarithmic transformation $G(x) = \log(1+rx)/r$, $r > 0$, can be assumed. The PH and PO models are special cases in both transformation classes. Then the parameter $r$ in $G(\cdot)$ can be estimated together with other parameters in the likelihood, as we will demonstrate in our simulation studies.

\section{Estimation and Inference} \label{sec_estimation_test_procedure}
\subsection{Sieve likelihood with Bernstein polynomials} \label{subsec_sieve}
In our likelihood function, we are interested in estimating the unknown parameter $\theta \in \Theta$:
$$\Theta = \{ \theta = (\beta_1^T, \beta_2^T, \alpha, \kappa, \Lambda_{1}, \Lambda_{2})^T \in \mathcal{B} \otimes \mathcal{M} \otimes \mathcal{M} \}.$$
Here $\mathcal{B}=\{(\beta=(\beta_1^T,\beta_2^T)^T, \alpha,\kappa) \in R^p \times R^{(0,1]} \times R^{+},\Vert\beta\Vert + \Vert\alpha\Vert + \Vert\kappa\Vert \leq M\}$ with $p$ being the dimension of $\beta$ and $M$ being a positive constant. We denote by $\mathcal{M}$ the collection of all bounded, continuous and nondecreasing, nonnegative functions over $[c, u]$, where $0 \leq c < u < \infty$. In practice, $[c,u]$ can be chosen as the range of all $L_{ij}$ and $R_{ij}$.

In our log-likelihood function $l_{n}(\theta; D)=\log{L_n(\theta; D)}=\sum_{i=1}^{n}\log{L(\theta; D_i)} = \sum_{i=1}^{n}l(\theta; D_i)$, there are finite-dimensional parameters of interest $(\beta,\alpha,\kappa)$ and two infinite-dimensional nuisance parameters $(\Lambda_{1},\Lambda_{2})$, which need to be estimated simultaneously. Unlike the right-censored data, tools like partial likelihood and martingale can not be applied to the interval-censored data due to the absence of exact event times. Instead, following \citet{univariate_PO_sieve}, we employ the sieve approach and form a sieve likelihood. Specifically, similar to \citet{frailty_case_II_transformation_sieve}, we use Bernstein polynomials to build a sieve space $\Theta_n = \{ \theta_n = (\beta^T, \alpha, \kappa, \Lambda_{1n}, \Lambda_{2n})^T \in \mathcal{B} \otimes \mathcal{M}_{n} \otimes \mathcal{M}_{n} \}$. Here, $\mathcal{M}_n$ is the space defined by Bernstein polynomials:
$$\mathcal{M}_n = \biggl\{ \Lambda_{jn}(t) = \sum_{k=0}^{m_{n}} \phi_{jk}B_{k}(t,m_{n},c,u): \sum_{k=0}^{m_n} |\phi_{jk}| \leq M_{n}; \ 0 \leq \phi_{j0} \leq \cdot \cdot \cdot \leq \phi_{jm_{n}}; j=1,2 \biggl\},$$
where $B_{k}(t,m_{n},c,u)$ represents the Bernstein basis polynomial defined as:
\begin{equation}
B_{k}(t,m_{n},c,u) =  {m_{n} \choose k} (\frac{t-c}{u-c})^{k} (1-\frac{t-c}{u-c})^{m_{n}-k}; \ k = 0,...,m_{n}, \label{Bern}
\end{equation}
with degree $m_{n} = o(n^{\nu})$ for some $\nu \in (0,1)$. We assume the basis polynomials $B_{k}(t,m_{n},c,u)$ are the same between the two margins, while the coefficients $\phi_{jk}$ can be margin-specific. In practice, one may choose $m_n$ based on model AIC values. With a pre-specified $m_n$, we solve $\phi_{jk}$ together with other parameters $(\beta, \alpha, \kappa)$. One big advantage of Bernstein polynomials is that they can achieve the non-negativity and monotonicity properties of $\Lambda_j(t)$ through re-parameterization \citep{frailty_case_II_transformation_sieve}. Another advantage of Bernstein polynomials is that they do not require the specification of interior knots, as seen from (\ref{Bern}), making them flexible for use.

With the sieve space defined above, $\Lambda_j(t)$  will be approximated by $\Lambda_{jn}(t) \in \mathcal{M}_n$. In the next section, we propose an estimation procedure to maximize $l_n(\theta; D)$ over the sieve space $\Theta_n$ to obtain the sieve maximum likelihood estimators $\hat{\theta}_n = (\hat{\beta}_n^T, \hat{\alpha}_n, \hat{\kappa}_n, \hat{\Lambda}_{1n}, \hat{\Lambda}_{2n})^T$.

\subsection{Estimation procedure for sieve maximum likelihood estimators $\hat{\theta}_n$} \label{subsec_two_step}
We develop a novel sieve maximum likelihood estimation procedure that is generally applicable to any choice of Archimedean copulas and marginal models. In principle, we can obtain the sieve maximum likelihood estimators by maximizing the joint likelihood function (\ref{joint_likelihood}) in one step. Due to the complex structure of the joint likelihood function, we recommend using a separate step to obtain appropriate initial values for all the unknown parameters. In essence, $(\beta_j, \Lambda_{jn})$ are first estimated marginally in step 1(a). Then their estimators are plugged into the joint likelihood to form a pseudo-likelihood. In step 1(b), the dependence parameters $(\alpha, \kappa)$ are estimated through maximizing the pseudo-likelihood function. Finally, using initial values from step 1(a) and 1(b), we update all the unknown parameters simultaneously under the joint log-likelihood function in step 2. The estimation procedure is described below:
\begin{enumerate}
\item Obtain initial estimates of $\theta_n$:
\begin{enumerate}
\item $(\hat{\beta}_{jn}^{(1)}, \hat{\Lambda}_{jn}^{(1)}) = \argmaxC_{(\beta_j, \Lambda_{jn})} l_{jn}(\beta_j, \Lambda_{jn})$, where $l_{jn}$ denotes the sieve log-likelihood for the marginal model, $j=1, 2$;
\vspace{0.1cm}
\item $(\hat{\alpha}_{n}^{(1)},\hat{\kappa}_{n}^{(1)})=\argmaxC_{(\alpha,\kappa)} l_n(\hat{\beta}_{n}^{(1)}=(\hat{\beta}_{1n}^{(1)},\hat{\beta}_{2n}^{(1)}), \alpha, \kappa, \hat{\Lambda}_{1n}^{(1)},\hat{\Lambda}_{2n}^{(1)})$, where $\hat{\beta}_{jn}^{(1)}$ and $\hat{\Lambda}_{jn}^{(1)}$ are the initial estimates from (a), and $l_n$ is the joint sieve log-likelihood.
\end{enumerate}

\item Simultaneously maximize the joint sieve log-likelihood to get final estimates:\\
$\hat{\theta}_n = (\hat{\beta}_n,\hat{\alpha}_n,\hat{\kappa}_n,\hat{\Lambda}_{1n},\hat{\Lambda}_{2n})=\argmaxC_{(\beta,\alpha,\kappa,\Lambda_{1n},\Lambda_{2n})} l_n(\beta,\alpha,\kappa,\Lambda_{1n},\Lambda_{2n})$ with initial values $(\hat{\beta}_{n}^{(1)},\hat{\alpha}_{n}^{(1)},\hat{\kappa}_{n}^{(1)},\hat{\Lambda}_{1n}^{(1)},\hat{\Lambda}_{2n}^{(1)})$ obtained from step 1(a) and 1(b).
\end{enumerate}

For the variance-covariance of finite-dimensional parameter estimates ($\hat{\beta}_n,\hat{\alpha}_n,\hat{\kappa}_n$), we invert the observed information matrix of all parameters including the nuisance parameters ($\phi_{jk}$) from the last iteration of step 2 and then take the corresponding block. In section \ref{subsec_asymptotics}, we establish the asymptotic normality and semiparametric efficiency for the finite-dimensional parameters. However, since the asymptotic variance form is intractable, we adopt this heuristic approach, which has been shown to work well in practice \citep{ding2011}.

Some standard optimization algorithms such as the Newton-Raphson algorithm or the conjugate gradient algorithm can be employed to obtain the maximizers and observed information matrix. Due to the complex structure of the joint sieve log-likelihood, instead of analytically deriving the first and second order derivatives, we propose to use the Richardson's extrapolation (\citeauthor{Richardson}, \citeyear{Richardson}) to approximate the score function and observed information matrix numerically. As shown in our simulations, the proposed procedure guarantees almost $100\%$ convergence and the computing speed is notably improved by using initial values from step 1.

\subsection{Asymptotic properties of sieve estimators} \label{subsec_asymptotics}
This section presents asymptotic properties of the sieve maximum likelihood estimators $\hat{\theta}_n$ with regularity conditions and proofs being supplied in Appendix D of the Supplementary Materials. Denote $P$ as the true probability measure and $\mathbb{P}_n$ as the empirical measure for $n$ independent subjects. Let $\vert v\vert$ be the Euclidean norm for a vector $v$. Define the supremum norm $\Vert f \Vert_{\infty} = sup_t\vert f(t)\vert$ for a function $f(t)$. Also define $\Vert f \Vert_{L_2(P)} = (\int \vert f \vert^2 dP)^{1/2}$ for a function $f$ under the probability measure $P$. In particular, the $L_2(P)$ norm for $\Lambda_j$ is defined as $\Vert \Lambda_j \Vert_2^2 = \int [\{\Lambda_j(l)\}^2 +  \{\Lambda_j(r)\}^2 ] dF_j(l,r)$, where $F_j(l,r)$ denotes the joint cumulative distribution function of $L_{ij}$ and $R_{ij} \ (i=1,...,n; j = 1,2)$. Finally, we define the distance between $\theta_1 = (\beta_1^T,\alpha_1,\kappa_1,\Lambda_{11},\Lambda_{21})^T \in \Theta$ and $\theta_2 = (\beta_2^T,\alpha_2,\kappa_2,\Lambda_{12},\Lambda_{22})^T \in \Theta$ as
$$ d(\theta_1,\theta_2) = ( \vert \beta_1-\beta_2 \vert^2 + \vert \alpha_1-\alpha_2 \vert^2 + \vert \kappa_1-\kappa_2 \vert^2 + \Vert \Lambda_{11}-\Lambda_{12}\Vert_{2}^2 + \Vert \Lambda_{21}-\Lambda_{22} \Vert_{2}^2 )^{1/2}.$$
Let $\theta_0 = (\beta_0^T,\alpha_0,\kappa_0,\Lambda_{10},\Lambda_{20})^T$ denote the true value of $\theta \in \Theta$. The following theorems present the convergence rate, asymptotic normality, and efficiency of the sieve estimators.

\begin{theorem} \label{thm:rate}
(Convergence rate) Assume that Conditions 1-2 and 4-5 in Appendix D of the Supplementary Materials hold. Let $m_n = o(n^{\nu})$, where $\nu \in (0,1)$ and $q$ be the smoothness parameter of $\Lambda_j$ as defined in Condition 4, then we have
$$d(\hat{\theta}_n,\theta_0) = O_p\big( n^{-\min\{{q\nu}/{2}, (1-\nu)/{2}\}} \big).$$
\end{theorem}
Theorem \ref{thm:rate} states that the sieve estimator $\hat{\theta}_n$ has a polynomial convergence rate. Although this overall convergence rate is lower than $n^{-1/2}$, the following normality theorem states that the proposed estimators of the finite-dimensional parameters ($\beta, \alpha, \kappa$) are asymptotically normal and semiparametrically efficient.

\begin{theorem} \label{thm:normality}
(Asymptotic normality and efficiency) Suppose Conditions 1-5 in Appendix D hold. Define $\hat{b}_n = (\hat{\beta}_n^T, \hat{\alpha}_n, \hat{\kappa}_n)^T$ and $b_0 = (\beta_0^T, \alpha_0, \kappa_0)^T$. If $1/(2+q)< \nu < {1}/{2}$, then
$${n}^{1/2}(\hat{b}_n-b_0) = I^{-1}(b_0) n^{1/2} \mathbb{P}_n l^{*}(b_0, \Lambda_{10}, \Lambda_{20}; D) + o_p(1) \to_d N\{0, I^{-1}(b_0)\},$$
where $I(b_0) = Pl^{*}(b_0, \Lambda_{10}, \Lambda_{20}; D)^{\otimes 2}$ and $l^{*}(b_0, \Lambda_{10}, \Lambda_{20}; D)$ is the efficient score function defined in the proof. Therefore, $\hat{b}_n$ is asymptotically normal and efficient.
\end{theorem}

\subsection{Generalized score test} \label{subsec_score}
We now separate $\beta$ into two parts: $\beta_g$ and $\beta_{ng}$, where $\beta_g$ is the parameter set of interest for hypothesis testing and $\beta_{ng}$ denotes the rest of the regression coefficients. The likelihood-based tests such as the Wald, score, and likelihood-ratio tests can be constructed to test $\beta_g$, and they are asymptotically equivalent. In our motivating study, we aim to perform a GWAS on AMD progression, which contains testing millions of SNPs one-by-one. Therefore, computing speed is critical. We propose to use the generalized score test. One big advantage of the score test in a GWAS setting is, one only needs to estimate the unknown parameters once under the null model without any SNP (i.e., $\beta_g = 0$), since the non-genetic risk factors are the same no matter which SNP is being tested. Therefore, the score test is faster as compared to the Wald and likelihood ratio tests. Moreover, the Wald or likelihood ratio test needs to estimate parameters under each alternative hypothesis (a total of 6 millions in our real data application), which may fail when the estimation procedure fails to converge.

With the sieve joint likelihood, we can obtain the restricted sieve maximum likelihood estimators under $H_0$  ($\beta_g = 0$ and the rest parameters are arbitrary), and then calculate the generalized score test statistic as defined in \citet{theoretical_stat}. Similar to our estimation procedure, we also propose to use Richardson's extrapolation to numerically approximate the first and second order derivatives when calculating the score test statistic.

\section{Simulation study} \label{sec_simulations}
We first evaluated the parameter estimation of our proposed two-parameter copula sieve model (i.e., its transformation margins are approximated by sieves) for bivariate data under general interval censoring. Then we assessed the type-I error control, and power performance of the proposed generalized score test. We also evaluated the accuracy in estimating the joint survival probability using our proposed method. Finally, we evaluated the computing speed and convergence rate of our proposed method.

\subsection{Generating bivariate interval-censored times} \label{generating data}
The data were generated from various Archimedean copula models (i.e., Clayton, Frank, Ali--Mikhail--Hap (AMH) and Joe) with Loglogistic margins. We first generated bivariate true event times $T_{ij}$ using the conditioning approach described in \citet{Sun_LIDA_2018}. To obtain interval-censored data, we followed the censoring procedure in \citet{simulate_IC}, which fits the study design of AREDS data. Explicitly, we assumed each subject was assessed for $K$ times with the length between two adjacent assessment times following an Exponential distribution. In the end, for each subject $i$, $L_{ij}$ was defined as the last assessment time before $T_{ij}$ and $R_{ij}$ was the first assessment time after $T_{ij}$. The overall right-censoring rate is set to be $25\%$. For the dependence strength between margins, we set Kendall's $\tau$ at 0.2 or 0.6, indicating weak or strong dependence. We assumed that the two event times share a common baseline distribution, for example, PO model with Loglogistic baseline hazards function (scale $\lambda=1$ and shape $k=2$) or PH model with Weibull baseline hazards function (scale $\lambda=0.1$ and shape $k=2$).

We included both genetic and non-genetic covariates in the simulations. Specifically, each SNP, coded as 0 or 1 or 2, was generated from a multinomial distribution with probabilities $\{(1-p)^2,2p(1-p),p^2\}$, where $p = 40\%$ or $5\%$ is the minor allele frequency (MAF). We also included a margin-specific continuous variable, generated from $N(6,2^2)$, and a subject-specific binary variable, generated from Bernoulli ($p=0.5$).

Under all scenarios, the sample size was set as $N=500$. For simplicity, we assumed the same covariate effects for two margins, denoted as $(\beta_{ng1},\beta_{ng2},\beta_g)$, where $\beta_{ng1}$ and $\beta_{ng2}$ are marginal- and subject-specific non-genetic effects, respectively, and $\beta_g$ is the SNP effect. We set $\beta_{ng1} = \beta_{ng2}=0.1$. For estimation performance evaluation, we let $\beta_g=0$ and replicated 1,000 times. For type-I error control evaluation of testing $\beta_g = 0$, we replicated 100,000 times and evaluated at various tail levels: 0.05, 0.01, 0.001 and 0.0001, respectively. For power evaluation, we replicated 1,000 times under each SNP effect size, where a range of $\beta_g$'s were selected to represent weak to strong SNP effects.

\subsection{Simulation-I: parameter estimation} \label{Estimation}
In this section, we evaluated the estimation performance of our proposed method under both correct and misspecified settings. For the margins, we used the true linear transformation function. We assumed the same Bernstein coefficients $\phi_{1k}=\phi_{2k}$ with degree $m_n=3$ ($k=0,1,2,3$) for both $\Lambda_1$ and $\Lambda_2$. For the event time range $[c,u]$, we chose $c=0$ and set $u$ as the largest value of all $\{L_{ij}, R_{ij}\}$ plus a constant.

In Table \ref{tab:Estimation_case_II}(a), the true model is Clayton copula with Loglogistic (PO) or Weibull (PH) margins, under Kendall's $\tau = 0.6$. We fitted three models: the true copula model with parametric margins (i.e., Clayton copula with Loglogistic or Weibull margins, denoted as ``Clayton-PM''), a two-parameter copula sieve model  (``Copula2-S'') and a marginal sieve model (i.e., the marginal transformation model approximated by sieves) where the variance-covariance is estimated by the robust sandwich estimator (``Marginal-S'') (a model also used in \citeauthor{frailty_case_II_transformation_sieve}, \citeyear{frailty_case_II_transformation_sieve}). We obtained estimation biases and 95\% coverage probabilities for regression coefficients and dependence parameters. Under the two-parameter copula model, the sieve maximum likelihood estimators are all virtually unbiased, and all empirical coverage probabilities are close to the nominal level. Moreover, their standard errors are virtually the same as the standard errors under the true parametric model, indicating our proposed method works well. For the robust marginal sieve model, the regression coefficient estimates are also unbiased with correct coverage probabilities, but their standard errors are apparently larger.

We further evaluated the estimation performance of the proposed model on bivariate interval-censored data generated from copula models that do not belong to the two-parameter copula family, such as Frank copula with $\tau = 0.6$, AMH copula with $\tau = 0.2$ ($\tau$ is always $<\frac{1}{3}$ for AMH copula) and Joe copula with $\tau = 0.6$. In Table \ref{tab:Estimation_case_II}(b), the regression coefficient estimates from the two-parameter copula are all unbiased with coverage probabilities close to 95\%. The biases for the $\tau$ estimates are also minimal with good coverage probabilities except in the scenario when data were generated from a Joe copula (coverage probability = 83\%). Overall, the two-parameter copula model family demonstrates good robustness to misspecification in copula functions.

In the real setting, the value of the transformation function parameter $r$ is often unknown. Therefore, we examined our methods in estimating the transformation function parameter $r$ together with the other parameters in our proposed model. The results are presented in the Table 1 of Appendix A in the Supplementary Materials, which shows satisfactory estimation performance for all parameters including the transformation parameter.

\subsection{Simulation-II, generalized score test performance} \label{test_performance}
We evaluated the type-I error control of our proposed generalized score test under Copula2-S. Specifically, we tested the SNP effect $\beta_g$ under different dependence strengths (Kendall's $\tau=0.6, \ 0.2$) and two different MAFs (40\%, 5\%). The true model is Clayton copula with Loglogistic margins. We included score tests of two misspecified copula models, one with misspecified margins but correct copula (i.e., Clayton copula with Weibull margins) and the other with misspecified copula but correct margins (i.e., Gumbel copula with Loglogistic margins). We also included the score test under the correct parametric copula model (i.e., Clayton copula with Loglogistic margins), which served as the benchmark model. Besides, we examined Wald tests from the marginal Loglogistic model with variance-covariance either from the independence estimate (i.e., the naive estimate assuming two margins are independent) or the robust sandwich estimate (i.e., accounting for the correlation between two margins).

Table \ref{tab:Type_I_error} shows type I errors under different tail levels. In the top part where Kendall's $\tau=0.6$, our proposed score test controls type-I errors as well as the correct parametric model at all tail levels and MAFs. However, type-I errors in the two misspecified copula models are inflated at all scenarios, especially when margins are wrong at MAF $ = 40\%$. It is not surprising to observe the greatest inflation occurs with the marginal approach under the independence assumption. After applying the robust variance-covariance estimate, the type-I errors are controlled at MAF = 40\% but still slightly inflated at MAF = 5\%. When Kendall's $\tau=0.2$, the proposed two-parameter model still performs as well as the correct parametric model and outperforms the other models, although the type-I error inflations from other models were smaller due to the weaker dependence.

\begin{figure}[!p]
\centering\includegraphics[scale=0.43]{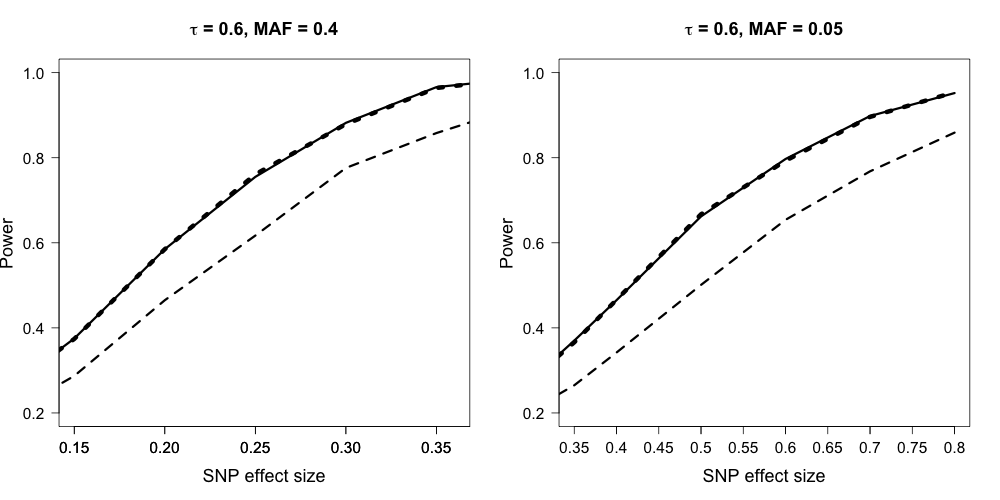}
\caption{Simulation results for power performance of the score test under three models: Clayton-LL (top dashed curve), Copula2-S (solid curve) and Marginal-S (bottom dashed curve).}
\label{fig:Clayton_Loglog_power}
\end{figure}

We also compared the power performance between the score test under our Copula2-S model and score tests from two other models: the true parametric copula model and the Marginal-S model. Figure \ref{fig:Clayton_Loglog_power} presents the power curves of these three tests over a range of SNP effect sizes. Our proposed model yields the similar power performance as the true parametric model and is considerably more potent than the robust marginal sieve model.

\subsection{Simulation-III: joint survival probability estimation performance}
In addition, we evaluated the accuracy for estimating joint survival probabilities under our proposed Copula2-S model. We generated data from the Clayton copula with Weibull margins, and fitted the Clayton-Weibull (``Clayton-WB'') and Copula2-S models and obtained the average estimated joint survival probabilities $Pr(T_1 > t, T_2 > t|Z_1, Z_2)$ on a sequence of pre-specified time points given covariate values. The number of replications is $1,000$. In Appendix A of the Supplementary Materials, Figure S1 illustrates that Copula2-S produced an almost identical joint survival profile as Clayton-WB. In addition, we quantified the estimation error between the estimated and true joint survival probabilities by the mean square errors (MSE) averaged over all time points and replications, which are $0.0004$ (sd $=0.0012$) and $0.0003$ (sd $=0.0005$) for Copula2-S and Clayton-WB, respectively, indicating the probabilities are well estimated.

\subsection{Simulation-IV, convergence and computing speed} \label{Speed}
We examined the computational advantages of our proposed two-step sieve estimation procedure as compared to the one-step estimation approach (i.e., directly maximizes the joint likelihood with arbitrary initial values). Data were simulated from a Clayton copula with Loglogistic margins. For $1,000$ replications, the one-step procedure took $1,260$ seconds while our proposed procedure took $925$ seconds, saving about $27\%$ computing time. For convergence rate, the proposed procedure failed in $0.1\%$ times, whereas the one-step procedure failed in $1.6\%$ times.

We also compared the computing speed of three likelihood-based tests on testing $1,000$ SNPs under three models: the true Clayton model with Loglogistic margins, our proposed Copula2-S model and the Marginal-S model. The 1,000 genetic variants were simulated from MAF $=40\%$. The results are shown in Table S3 in the Appendix A from the Supplementary Materials. We found that the score test is about 3-5 times faster than the Wald test or the likelihood ratio test on average. Within the three score tests, although the score test under our Copula2-S model is the slowest due to model complexity, it is still faster than the Wald test under the Marginal-S model. Given its advantages in robustness, type-I error control, and power performance, we recommend the proposed Copula2-S model with the score test for the large-scale testing case.

\section{Real data analysis} \label{sec_real_data}
We implemented our proposed method to analyze the AREDS data. AREDS was designed to assess the clinical course of, and risk factors for the development and progression of AMD. DNA samples were collected from the consenting participants and genotyped by the International AMD Genomics Consortium \citep{AMD_genetic_2016}. In this study, each participant was examined every six months in the first six years and then every year after year six. To measure the disease progression, a severity score, scaled from one to twelve (with a more significant value indicating more severe AMD), was determined for each eye of each participant at every examination. The outcome of interest is the bivariate progression time-to-late-AMD, where late-AMD is defined as the stage with severity score $\ge 9$. Both phenotype and genotype data of AREDS are available from the online repository dbGap (accession: phs000001.v3.p1, and phs001039.v1.p1, respectively). By far, all the studies that analyzed the AREDS data for AMD progression treated the outcome as right-censored (e.g., \citet{AMD_prog_3}, \citet{Yan_2018}, and \citet{Sun_LIDA_2018}), and some only used data from the worst eye in each subject (e.g., \citet{Seddon_nine}).

We analyzed 2,718 Caucasian participants, including 2,295 subjects who were free of late-AMD in both eyes at the enrollment, i.e., time $0$ (bivariate data indicated as group A), and 423 subjects who had one eye already progressed to late-AMD by enrollment (univariate data indicated as group B). For the $j$th eye (free of late-AMD at time 0) of subject $i$, we observe $L_{ij}$, the last assessment time when the $j$th eye was still free of late-AMD and $R_{ij}$, the first assessment time when the $j$th eye was already diagnosed as late-AMD. For the eye that did not progress to late-AMD by the end of the study follow-up, we assigned a large number to $R_{ij}$. Since there are two groups of subjects (group A and B), we implemented a two-part model. Specifically, we created a covariate for each eye to indicate whether its fellow eye had already progressed or not at time 0 (i.e., $0$ for both eyes of group A subjects and $1$ for group B subjects). Then the joint likelihood is the product of the copula sieve model for group A subjects and the marginal sieve model for group B subjects. In addition, we performed a secondary sensitivity analysis using only group A subjects and obtained similar top SNPs as from the two-part model (Table S4 of Appendix B in the Supplementary Materials).

We included four risk factors as non-genetic covariates, including the baseline age, severity score, smoking status, and fellow-eye progression status. We checked various combinations of transformation functions (i.e., $g(x) =x$ for PH model and $g(x) = \log (1+x)$ for PO model) and Bernstein polynomial degrees $m_n$ (from 3 to 6). We chose the model that produced the smallest $\textsc{aic}$, which is the PO model (i.e., $g(x) = \log (1+x)$) with $m_n = 4$ for both margins. The $\textsc{aic}$ results are summarized in Table S5 of Appendix B in the Supplementary Materials.

We performed GWAS on 6 million SNPs (either from exome chip or imputed) with MAF $> 5\%$ across the 22 autosomal chromosomes and plotted their $-\log(p)$ in Figure S2 in the Appendix B of the Supplementary Materials. As highlighted in the figure, the \textit{PLEKHA1--ARMS2--HTRA1} region on chromosome 10 and the \textit{CFH} region on chromosome 1 have variants reaching the ``genome-wide" significance level ($p< 5\times 10^{-8}$). Previously, these two regions were found being significantly associated with AMD onset from multiple case-control studies \citep{AMD_genetic_2016}. Moreover, we identified SNPs in a previously unrecognized \textit{ATF7IP2} region on chromosome 16, showing moderate to strong association with AMD progression ($ 5 \times 10^{-8} < p < 1 \times 10^{-5}$). As a comparison, we also fitted the robust marginal sieve model (Marginal-S) and the Gamma frailty sieve model (Frailty-S) \citep{frailty_case_II_transformation_sieve}, and performed the corresponding score tests for each SNP. Overall, their results are consistent with our Copula2-S model, but the $p$-values are generally larger (as shown in Table \ref{tab:GWAS_table}). Note that the \textit{CFH} region did not reach the ``genome-wide" significance level under the Marginal-S model.

Table \ref{tab:GWAS_table} presents the top significant variants of the three regions denoted in Figure S2. Besides Copula2-S, we also present score test $p$-values from Frailty-S and Marginal-S. The odds ratio of an SNP was calculated by fitting a Copula2-S model including this SNP and those non-genetic factors. For example, $rs2284665$, a known AMD risk variant from \textit{HTRA1} region, has an estimated odds ratio of 1.66 (95\% CI $=[1.46, 1.89]$), which implies its minor allele has a ``harmful'' effect on AMD progression. Under this model, the estimated dependence parameters are $\hat{\alpha} = 0.90$ and $\hat{\kappa} = 1.00$, corresponding to $\hat{\tau} = 0.40$, which indicates moderate dependence in AMD progression between two eyes.

\begin{figure}[!p]
\centering\includegraphics[scale=0.55]{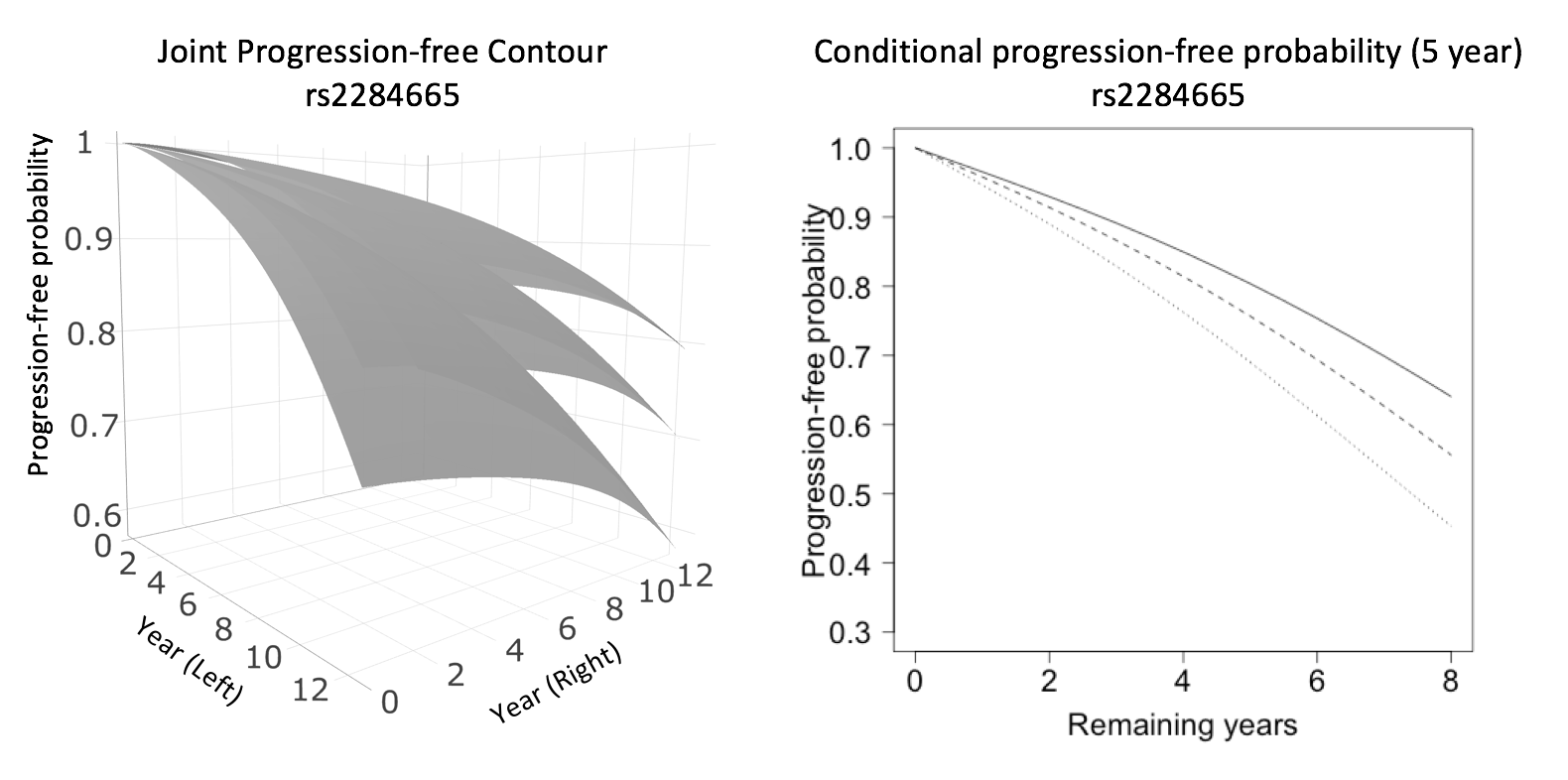}
\caption{Estimated progression-free probabilities for subjects with different genotypes of $rs2284665$ (smokers with age 68.6 and severity score 3.0 in both eyes). Left: joint progression-free probability contours (from top to bottom: $GG, GT, TT$); Right: conditional progression-free probability of remaining years (after year 5) for one eye, given the other eye has progressed by year 5 (from top to bottom: $GG, GT, TT$).}
\label{fig:3D_and_conditional}
\end{figure}

For variant $rs2284665$, we obtained both estimated joint and conditional survival functions from the fitted Copula2-S model. The left panel of Figure \ref{fig:3D_and_conditional} plots the joint progression-free probability contours for subjects who are smokers with the same age (= 68.6) and AMD severity score (= 3.0 for both eyes) but different genotypes of $rs2284665$. The right panel of Figure \ref{fig:3D_and_conditional} plots the corresponding conditional progression-free probability of remaining years (after year 5) for one eye, given its fellow eye has progressed by year 5. It is clearly seen that in both plots, the three genotype groups are well separated, with the $GG$ group having the largest progression-free probabilities. These estimated progression-free probabilities provide valuable information to characterize or predict the progression profiles for AMD patients with different characteristics.

\section{Conclusion and Discussion} \label{sec_conclusions}
We proposed a flexible copula-based semiparametric transformation model for analyzing and testing bivariate (general) interval-censored data. Unlike the approach proposed by \citet{HuCaseICopulaPH_2017}, which approximated the copula function by sieves, our approach kept the copula function in its parametric form but flexibly modeled the margins through semiparametric transformation models. In this way, our method guaranteed to produce consistent estimates for both regression and copula parameters, which then led to reliable joint distribution estimates. On the other hand, \citet{HuCaseICopulaPH_2017} focused on estimating regression parameters only but with possible biased estimates for the copula function. Our proposed method has the great advantage in computation and it is applicable to analyze large data sets and to perform a large number of tests. All the methods have been built into an R package \{CopulaCenR\}, which includes a variety of copula functions (e.g., Copula2, Clayton, Gumbel, Frank, Joe, AMH) and is available on CRAN at {https://cran.r-project.org/package=CopulaCenR}. The key R codes for this article can be found at https://github.com/yingding99/Copula2S.

Several model selection procedures have been proposed for copula-based methods. For example, the AIC is widely used for model selection purpose in copula models. \citet{wang_selection_2000} proposed a model selection procedure based on the nonparametric estimation of the bivariate joint survival function within Archimedean copulas. For model diagnostics, \citet{Chen_selection_2010} proposed a penalized pseudo-likelihood ratio test for copula models in complete data. Recently, \citet{Zhang2016_PIOS} developed a goodness-of-fit test for copula models using the pseudo in-and-out-of-sample method. To the best of our knowledge, there is no existing goodness-of-fit test for copula models of bivariate interval-censored data. In our real data analysis, we used AIC to guide the model selection for simplicity. However, a formal test for goodness-of-fit is desirable, especially for bivariate interval-censored data under the regression setting. It is worthwhile to investigate it as a future research direction.

We applied our method to a GWAS of AMD progression and successfully identified variants from two known AMD risk regions (\textit{CFH} on chromosome 1 and \textit{PLEKHA1--ARMS2--HTRA1} on chromosome 10) being significantly associated with AMD progression. Moreover, we also discovered variants from a region (\textit{ATF7IP2} on chromosome 16), which has not been reported before, showing moderate to strong association with AMD progression. On the contrary, we found that some known AMD risk loci (e.g., $rs12357257$ from \textit{ARHGAP21} on chromosome 10, $p=0.12$) are not associated with AMD progression. Therefore, the variants associated with risks of having AMD may not be necessarily associated with the disease progression; while some variants may be only associated with AMD progression but not with the disease onset. Our work is the first research on AMD progression which adopts a solid statistical model that appropriately handles bivariate interval-censored data. Our findings provided new insights into the genetic causes on AMD progression, which are critical for establishing novel and reliable predictive models of AMD progression to identify high-risk patients at an early stage accurately. Our proposed method applies to general bilateral diseases and complex diseases with co-primary endpoints.

\section*{Supplementary Materials} \label{Appendix}
Supplementary materials are available online at
%\href{http://biostatistics.oxfordjournals.org}%
{http://biostatistics.oxfordjournals.org}.

\bibliographystyle{biorefs}
\bibliography{refs_TS}

\begin{thebibliography}{99}

\bibitem[Chen \emph{and others}(2014)Chen, Chen, Lin and
  Tong]{chen2014analysis}
\textsc{Chen, M., Chen, L., Lin, K. and Tong, X.} (2014).
\newblock Analysis of multivariate interval censoring by diabetic retinopathy
  study.
\newblock {\em Communications in Statistics-Simulation and
  Computation\/}~\textbf{43}(7), 1825--1835.

\bibitem[Chen \emph{and others}(2007)Chen, Tong and Sun]{chen2007proportional}
\textsc{Chen, M., Tong, X. and Sun, J.} (2007).
\newblock The proportional odds model for multivariate interval-censored
  failure time data.
\newblock {\em Statistics in medicine\/}~\textbf{26}(28), 5147--5161.

\bibitem[Chen \emph{and others}(2009)Chen, Tong and Sun]{chen2009frailty}
\textsc{Chen, M., Tong, X. and Sun, J.} (2009).
\newblock A frailty model approach for regression analysis of multivariate
  current status data.
\newblock {\em Statistics in medicine\/}~\textbf{28}(27), 3424--3436.

\bibitem[Chen \emph{and others}(2013)Chen, Tong and Zhu]{chen2013linear}
\textsc{Chen, M., Tong, X. and Zhu, L.} (2013).
\newblock A linear transformation model for multivariate interval-censored
  failure time data.
\newblock {\em Canadian Journal of Statistics\/}~\textbf{41}(2), 275--290.

\bibitem[Chen \emph{and others}(2010)Chen, Fan, Pouzo and
  Ying]{Chen_selection_2010}
\textsc{Chen, X., Fan, Y., Pouzo, D. and Ying, Z.} (2010).
\newblock Estimation and model selection of semiparametric multivariate
  survival functions under general censorship.
\newblock {\em Journal of Econometrics\/}~\textbf{157(2)}, 129--142.

\bibitem[Clayton(1978)Clayton]{Clayton}
\textsc{Clayton, D.~G.} (1978).
\newblock A model for association in bivariate life tables and its application
  in epidemiological studies of familial tendency in chronic disease incidence.
\newblock {\em Biometrika\/}~\textbf{65}(1), 141--151.

\bibitem[Cook and Tolusso(2009)Cook and Tolusso]{copula_case_I_ph_piecewise}
\textsc{Cook, R.~J. and Tolusso, D.} (2009).
\newblock Second-order estimating equations for the analysis of clustered
  current status data.
\newblock {\em Biostatistics\/}~\textbf{10}(4), 756--772.

\bibitem[Cox and Hinkley(1979)Cox and Hinkley]{theoretical_stat}
\textsc{Cox, D.~R. and Hinkley, D.~V.} (1979).
\newblock {\em Theoretical Statistics\/}. Chapman \& Hall/CRC, London.

\bibitem[Ding \emph{and others}(2017)Ding, Liu, Yan  et~al.]{AMD_prog_3}
\textsc{Ding, Y., Liu, Y., Yan, Q.  \emph{and others}}. (2017).
\newblock Bivariate analysis of {Age-Related Macular Degeneration} progression
  using genetic risk scores.
\newblock {\em Genetics\/}~\textbf{206}(1), 119--133.

\bibitem[Ding and Nan(2011)Ding and Nan]{ding2011}
\textsc{Ding, Y. and Nan, B.} (2011).
\newblock A sieve {M}-theorem for bundled parameters in semiparametric models,
  with application to the efficient estimation in a linear model for censored
  data.
\newblock {\em Annals of Statistics\/}~\textbf{39}(1), 2795--3443.

\bibitem[Fritsche \emph{and others}(2016)Fritsche, IgI, Bailey
  et~al.]{AMD_genetic_2016}
\textsc{Fritsche, L.~G., IgI, W., Bailey, J.~N.  \emph{and others}}. (2016).
\newblock A large genome-wide association study of {Age-related Macular
  Degeneration} highlights contributions of rare and common variants.
\newblock {\em Nature Genetics\/}~\textbf{48}(2), 134--143.

\bibitem[Goethals \emph{and others}(2008)Goethals, Janssen and
  Duchateau]{copula_vs_frailty}
\textsc{Goethals, K., Janssen, P. and Duchateau, L.} (2008).
\newblock Frailty models and copulas: similarities and differences.
\newblock {\em Journal of Applied Statistics\/}~\textbf{35}(9), 1071--1079.

\bibitem[Goggins and Finkelstein(2000)Goggins and
  Finkelstein]{bivariate_cox_marginal}
\textsc{Goggins, W.~B. and Finkelstein, D.~M.} (2000).
\newblock A proportional hazards model for multivariate interval-censored
  failure time data.
\newblock {\em Biometrics\/}~\textbf{56}(3), 940--943.

\bibitem[Gumbel(1960)Gumbel]{gumbel}
\textsc{Gumbel, E.~J.} (1960).
\newblock Bivariate exponential distributions.
\newblock {\em Journal of the American Statistical
  Association\/}~\textbf{55}(292), 698--707.

\bibitem[Hu \emph{and others}(2017)Hu, Zhou and Sun]{HuCaseICopulaPH_2017}
\textsc{Hu, T., Zhou, Q. and Sun, J.} (2017).
\newblock Regression analysis of bivariate current status data under the
  proportional hazards model.
\newblock {\em Canadian Journal of Statistics\/}~\textbf{45}(4), 410--424.

\bibitem[Huang and Rossini(1997)Huang and Rossini]{univariate_PO_sieve}
\textsc{Huang, J. and Rossini, A.} (1997).
\newblock Sieve estimation for the proportional-odds failure-time regression
  model with interval censoring.
\newblock {\em Journal of the American Statistical
  Association\/}~\textbf{92}(439), 960--967.

\bibitem[Joe(1997)Joe]{Joe}
\textsc{Joe, H.} (1997).
\newblock {\em Multivariate Models and Dependence Concepts\/}. Chapman \& Hall,
  London.

\bibitem[Kiani and Arasan(2012)Kiani and Arasan]{simulate_IC}
\textsc{Kiani, K. and Arasan, J.} (2012).
\newblock Simulation of interval-censored data in medical and biological
  studies.
\newblock {\em International Journal of Modern Physics\/}~\textbf{9}, 112--118.

\bibitem[Kim and Xue(2002)Kim and Xue]{kim2002analysis}
\textsc{Kim, M.~Y. and Xue, X.} (2002).
\newblock The analysis of multivariate interval-censored survival data.
\newblock {\em Statistics in Medicine\/}~\textbf{21}(23), 3715--3726.

\bibitem[Kor \emph{and others}(2013)Kor, Cheng and Chen]{copula_case_II_ph_pw}
\textsc{Kor, C.~T., Cheng, K.~F. and Chen, Y.~H.} (2013).
\newblock A method for analyzing clustered interval-censored data based on
  {Cox} model.
\newblock {\em Statistics in Medicine\/}~\textbf{32}(5), 822--832.

\bibitem[Lindfield and Penny(1989)Lindfield and Penny]{Richardson}
\textsc{Lindfield, G.~R. and Penny, J. E.~T.} (1989).
\newblock {\em Microcomputers in Numerical Analysis\/}. Halsted Press, New
  York.

\bibitem[Nelsen(2006)Nelsen]{Nelson_2006}
\textsc{Nelsen, R.~B.} (2006).
\newblock {\em An Introduction to Copulas\/}. Springer-Verlag, New York.

\bibitem[Oakes(1982)Oakes]{frailty_oaks}
\textsc{Oakes, D.} (1982).
\newblock A model for association in bivariate survival data.
\newblock {\em Journal of the Royal Statistical Society: Series
  B\/}~\textbf{44}(3), 414--422.

\bibitem[Seddon \emph{and others}(2014)Seddon, Reynolds, Yu and
  Rosner]{Seddon_nine}
\textsc{Seddon, J., Reynolds, R., Yu, Y. and Rosner, B.} (2014).
\newblock Three new genetic loci are independently related to progression to
  advanced macular degeneration.
\newblock {\em PLoS ONE\/}~\textbf{9}(1), 1--11.

\bibitem[Sklar(1959)Sklar]{sklar}
\textsc{Sklar, A.} (1959).
\newblock Fonctions de r{\'e}partition {\`a} n dimensions et leurs marges.
\newblock {\em Publications de L'Institut de Statistique de L'Universit{\'e} de
  Paris\/}~\textbf{8}, 229--231.

\bibitem[Sun \emph{and others}(2019)Sun, Liu, Cook, Chen and
  Ding]{Sun_LIDA_2018}
\textsc{Sun, T., Liu, Y., Cook, R.~J., Chen, W. and Ding, Y.} (2019).
\newblock Copula-based score test for bivariate time-to-event data, with
  application to a genetic study of {AMD} progression.
\newblock {\em Lifetime Data Analysis\/}~\textbf{25}(3), 546--568.

\bibitem[Tong \emph{and others}(2008)Tong, Chen and Sun]{tong2008regression}
\textsc{Tong, X., Chen, M. and Sun, J.} (2008).
\newblock Regression analysis of multivariate interval-censored failure time
  data with application to tumorigenicity experiments.
\newblock {\em Biometrical Journal: Journal of Mathematical Methods in
  Biosciences\/}~\textbf{50}(3), 364--374.

\bibitem[Wang \emph{and others}(2008)Wang, Sun and Tong]{copula_case_I_ph}
\textsc{Wang, L., Sun, J. and Tong, X.} (2008).
\newblock Efficient estimation for the proportional hazards model with
  bivariate current status data.
\newblock {\em Lifetime Data Analysis\/}~\textbf{14}(2), 134--153.

\bibitem[Wang \emph{and others}(2015)Wang, Wang and
  McMahan]{wang2015regression}
\textsc{Wang, N., Wang, L. and McMahan, C.~S.} (2015).
\newblock Regression analysis of bivariate current status data under the
  gamma-frailty proportional hazards model using the {EM} algorithm.
\newblock {\em Computational Statistics \& Data Analysis\/}~\textbf{83},
  140--150.

\bibitem[Wang and Wells(2000)Wang and Wells]{wang_selection_2000}
\textsc{Wang, W. and Wells, M.~T.} (2000).
\newblock Model selection and semiparametric inference for bivariate
  failure-time data.
\newblock {\em Journal of the American Statistical
  Association\/}~\textbf{95}(449), 62--72.

\bibitem[Wen and Chen(2013)Wen and Chen]{wen_chen2013}
\textsc{Wen, C.~C. and Chen, Y.~H.} (2013).
\newblock A frailty model approach for regression analysis of bivariate
  interval-censored survival data.
\newblock {\em Statistica Sinica\/}~\textbf{23}(1), 383--408.

\bibitem[Wienke(2010)Wienke]{wienke2010frailty}
\textsc{Wienke, A.} (2010).
\newblock {\em Frailty models in survival analysis\/}. Chapman and Hall/CRC.

\bibitem[Yan \emph{and others}(2018)Yan, Ding, Liu  et~al.]{Yan_2018}
\textsc{Yan, Q., Ding, Y., Liu, Y.  \emph{and others}}. (2018).
\newblock Genome-wide analysis of disease progression in {Age-related Macular
  Degeneration}.
\newblock {\em Human Molecular Genetics\/}~\textbf{27}(5), 929--940.

\bibitem[Zeng \emph{and others}(2017)Zeng, Gao and
  Lin]{frailty_case_II_transformation_NPMLE}
\textsc{Zeng, D., Gao, F. and Lin, D.} (2017).
\newblock Maximum likelihood estimation for semiparametric regression models
  with multivariate interval-censored data.
\newblock {\em Biometrika\/}~\textbf{104}(3), 505--525.

\bibitem[Zhang \emph{and others}(2016)Zhang, Okhrin, Zhou and
  Song]{Zhang2016_PIOS}
\textsc{Zhang, S., Okhrin, O., Zhou, {Q. M.} and Song, {P. X. K.}} (2016).
\newblock Goodness-of-fit test for specification of semiparametric copula
  dependence models.
\newblock {\em Journal of Econometrics\/}~\textbf{193}(1), 215--233.

\bibitem[Zhou \emph{and others}(2017)Zhou, Hu and
  Sun]{frailty_case_II_transformation_sieve}
\textsc{Zhou, Q., Hu, T. and Sun, J.} (2017).
\newblock A sieve semiparametric maximum likelihood approach for regression
  analysis of bivariate interval-censored failure time data.
\newblock {\em Journal of the American Statistical
  Association\/}~\textbf{112}(518), 664--672.

\end{thebibliography}

\begin{table}[!p]
\caption{Estimation results for bivariate interval-censored data (a) fitted with three correctly-specified models: Clayton model with parametric margins (Loglogistic for proportional odds and Weibull for proportional hazards; denoted as Clayton-PM), two-parameter copula sieve model (Copula2-S) and marginal sieve model (Marginal-S); (b) fitted with the proposed Copula2-S model (misspecified copula) where the true data are generated from Frank, AMH, and Joe copulas.}
\label{tab:Estimation_case_II}
\resizebox{1.0\linewidth}{!}{
{\tabcolsep=4.25pt
\begin{tabular}{@{}cccccccccccc@{}}
\multicolumn{1}{c}{(a)}       &                 \multicolumn{11}{c}{}                                            \\
\tblhead{\multicolumn{1}{c}{}    &  \multicolumn{3}{c}{Clayton-PM}   &&  \multicolumn{3}{c}{Copula2-S}   && \multicolumn{3}{c}{Marginal-S}  \\
\cline{2-4}\cline{6-8}\cline{10-12}
\multicolumn{1}{c}{Param} &  Bias   &   SE   &    SEE (CP)  &&  Bias   &   SE   &    SEE (CP)  &&  Bias   &   SE   &    SEE (CP)
}
\multicolumn{1}{c}{}       &                 \multicolumn{11}{c}{{proportional odds}}                                            \\
\rule{0pt}{3ex}
$\beta_{ng1}$       & 0.0013  &  0.0171 &  0.0163 (0.942)  &&  0.0003  &   0.0176   &  0.0165 (0.938)  &&  0.0024  & 0.0293  &  0.0300 (0.930) \medskip \\
$\beta_{ng2}$       & 0.0120  &  0.1300 &  0.1300 (0.945)  &&  0.0006  &   0.1330   &  0.1310 (0.939)  &&  0.0110  & 0.1510  &  0.1500 (0.944) \medskip \\
$\beta_{g}$         & -0.0007 &  0.0927 &  0.0942 (0.953)  &&  -0.0110  &   0.0951   &  0.0947 (0.950)  &&  0.0012  & 0.1050  &  0.1090 (0.955) \medskip \\
$\tau$              & -0.0005 &  0.0210 &  0.0208 (0.944)  &&  -0.0045  &   0.0223   &  0.0221 (0.950)  &&  NA      & NA      &  NA      \medskip \\

\hline

\multicolumn{1}{c}{}       &                 \multicolumn{11}{c}{{proportional hazards}}                                            \\
\rule{0pt}{3ex}
$\beta_{ng1}$       & 0.0012    &  0.0097 &  0.0103 (0.958)   &&  0.0013     &   0.0099    &  0.0105 (0.957)   &&  0.0009    & 0.0182  &  0.0187 (0.957) \medskip \\
$\beta_{ng2}$      & -0.0043  &  0.0780 &  0.0789 (0.952)  &&  -0.0040   &   0.0782     &  0.0788 (0.951)   &&  -0.0043  & 0.0960  &  0.0969 (0.957) \medskip \\
$\beta_{g}$          & 0.0005    &  0.0606 &  0.0569 (0.935)  &&  0.0002    &   0.0608     &  0.0569 (0.938)  &&  0.0003    & 0.0722  &  0.0701 (0.945) \medskip \\
$\tau$                   & -0.0003  &  0.0220 &   0.0219 (0.952)  &&  -0.0012   &   0.0224      &  0.0221 (0.951)   &&  NA      & NA      &  NA
\lastline
\end{tabular}}
}

%\bigskip
\vspace{1cm}

\resizebox{1.0\linewidth}{!}{
{\tabcolsep=4.25pt
\begin{tabular}{@{}cccccccccccc@{}}
\multicolumn{1}{c}{(b)}       &                 \multicolumn{11}{c}{}                                            \\
\tblhead{\multicolumn{1}{c}{}    &  \multicolumn{3}{c}{Frank}    &&  \multicolumn{3}{c}{AMH}   && \multicolumn{3}{c}{Joe}  \\
\cline{2-4}\cline{6-8}\cline{10-12}
\multicolumn{1}{c}{Param} &  Bias   &   SE    &    SEE (CP)      &&  Bias     &   SE       &    SEE (CP)      &&  Bias    &   SE    &    SEE (CP)
}
$\beta_{ng1}$       & 0.0002   &  0.0177 &  0.0176 (0.950)  &&  -0.0011  &   0.0262   &  0.0267 (0.953)  && 0.0016 & 0.0160  &  0.0166 (0.962) \medskip \\
$\beta_{ng2}$       & 0.0018   &  0.1480 &  0.1470 (0.944)  &&  0.0013   &   0.1250   &  0.1250 (0.951)  &&  -0.0027  & 0.1388  &  0.1438 (0.954) \medskip \\
$\beta_{g}$         & 0.0001   &  0.1050 &  0.1060 (0.952)  &&  -0.0001 &   0.0885   &  0.0901 (0.959)  &&  0.0037  & 0.0984  &  0.1043 (0.962) \medskip \\
$\tau$              & -0.0036  &  0.0219 &  0.0198 (0.937)  &&  -0.0056  &   0.0318   &  0.0304 (0.934)  &&  0.0168  & 0.0195  &  0.0185 (0.830)
\lastline
\end{tabular}}
}

\end{table}

\begin{table}
\caption{Type-I error for the genetic effect $\beta_g$ at various tail levels. Six models were compared:  independent marginal Loglogistic model (Indep-LL), robust marginal Loglogistic model (Robust-LL), Clayton copula with Weibull margins (Clayton-W), Gumbel copula with Loglogistic margins (Gumbel-LL), two-parameter copula with transformation margins being approximated by sieves (Copula2-S) and the true Clayton copula and Loglogistic margins (Clayton-LL).}
\label{tab:Type_I_error}
{\tabcolsep=4.25pt
\begin{tabular}{@{}clllllll@{}}
\tblhead{\multicolumn{1}{l}{{MAF}} & {Tail}   & {Indep-LL}  & {Robust-LL}     & {Clayton-W} &  {Gumbel-LL}  & {Copula2-S}     & {Clayton-LL} }
\multicolumn{1}{l}{}             &                 & \multicolumn{5}{c}{{Kendall's $\tau=0.6$}}                                                                                                                                   \\
\multirow{4}{*}{{40\%}}   & {0.05}   & 0.141                       & 0.051                      & 0.131                & 0.065                       & 0.052                      & 0.050                \\
                                 & {0.01}   & 0.053                        & 0.010                     & 0.041                 & 0.015                       & 0.010                      & 0.010                 \\
                                 & {0.001}  & 0.0131                       & 0.0012                      & 0.0074                & 0.0022                        & 0.0013                      & 0.0012                \\
                                 & {0.0001} & 0.0037                       & 0.0002                      & 0.0012                & 0.0003                        & 0.0001                      & 0.0001                \\
\rule{0pt}{3ex}
\multirow{4}{*}{{5\%}}    & {0.05}   & 0.141                       & 0.056                      & 0.059                & 0.066                       & 0.053                       & 0.051                \\
                                 & {0.01}   & 0.053                       & 0.014                      & 0.012                & 0.016                       & 0.012                      & 0.011                \\
                                 & {0.001}  & 0.0136                       & 0.0018                      & 0.0013                & 0.0020                       & 0.0013                      & 0.0012                \\
                                 & {0.0001} & 0.0034                       & 0.0003                      & 0.0002                & 0.0003                       & 0.0002                      & 0.0002                \\

\hline

\multicolumn{1}{l}{}             &                 & \multicolumn{5}{c}{{Kendall's $\tau=0.2$}}                                                                                                                                    \\
\multirow{4}{*}{{40\%}}   & {0.05}   & 0.083                       & 0.051                      & 0.103                & 0.061                       & 0.051                      & 0.050                \\
                                 & {0.01}   & 0.022                       & 0.010                      & 0.029                & 0.013                       & 0.010                      & 0.010                \\
                                 & {0.001}  & 0.0036                       & 0.0012                      & 0.0045                & 0.0017                       & 0.0011                      & 0.0010                \\
                                 & {0.0001} & 0.0006                       & 0.0002                      & 0.0006                & 0.0003                       & 0.0002                     & 0.0002                \\
\rule{0pt}{3ex}
\multirow{4}{*}{{5\%}}    & {0.05}   & 0.083                       & 0.056                      & 0.054                & 0.060                       & 0.053                      & 0.052                \\
                                 & {0.01}   & 0.023                       & 0.013                      & 0.011                & 0.014                       & 0.012                      & 0.011                \\
                                 & {0.001}  & 0.0036                       & 0.0017                      & 0.0013                & 0.0018                       & 0.0014                      & 0.0013                \\
                                 & {0.0001} & 0.0007                       & 0.0003                       & 0.0001                & 0.0002                       & 0.0002                      & 0.0001
\lastline
\end{tabular}}
\end{table}

\begin{table}[!p]
\caption{The top SNPs identified to be associated with AMD progression. The last two columns come from the gamma frailty sieve model and the robust marginal sieve model, respectively.}
\label{tab:GWAS_table}
{\tabcolsep=4.25pt
\begin{tabular}{@{}llllllll@{}}
\tblhead{    SNP &  Chr & Gene & MAF & OR & $p$ (Copula2-S) & $p$ (Frailty-S) & $p$ (Marginal-S) }
    $rs2284665$ & 10 &  \textit{HTRA1} & 0.33 & 1.66 & $1.5 \times 10^{-14}$ & $2.7 \times 10^{-12}$ & $1.6 \times 10^{-10}$ \\
    $rs2293870$ &  10 &  \textit{ARMS2-HTRA1} & 0.33 & 1.65 & $2.5 \times 10^{-14}$  & $2.5 \times 10^{-12}$ & $ 2.4 \times 10^{-10}$ \\
    $rs3750846$ &  10 & \textit{ARMS2-HTRA1} & 0.34 & 1.62  & $1.6 \times 10^{-13}$ & $8.5 \times 10^{-12}$ & $ 8.7 \times 10^{-10}$  \\
    $rs58649964$  &  10 & \textit{PLEKHA1} & 0.24 & 1.63 & $3.0 \times 10^{-11}$  & $1.0 \times 10^{-9}$ & $ 2.0 \times 10^{-8}$ \\
    $rs10922109$ & 1 & \textit{CFH} & 0.28 & 0.64 & $4.0 \times 10^{-9}$ & $7.4 \times 10^{-9}$ & $7.4 \times 10^{-8} $ \\
    $rs1329427$ & 1 & \textit{CFH} & 0.28 & 0.64 & $4.4 \times 10^{-9}$ & $8.3 \times 10^{-9}$ & $ 8.1 \times 10^{-8} $ \\
    $rs10801559$ & 1 & \textit{CFH} & 0.28 & 0.64 & $4.8 \times 10^{-9}$ & $9.3 \times 10^{-9}$  & $ 8.8 \times 10^{-8} $ \\
    $rs1410996$ & 1 & \textit{CFH} & 0.28 & 0.64 & $5.3 \times 10^{-9}$  & $1.1 \times 10^{-8}$ & $ 1.0 \times 10^{-7} $ \\
    $rs12708701$ & 16 & \textit{ATF7IP2} & 0.13 & 1.62 & $1.1 \times 10^{-7}$ & $2.5 \times 10^{-7}$ & $ 7.0 \times 10^{-7} $  \\
    $rs28368872$ & 16 & \textit{ATF7IP2} & 0.13 & 1.62 & $1.3 \times 10^{-7}$  & $4.3 \times 10^{-7}$ & $ 8.7 \times 10^{-7} $
\lastline
\end{tabular}}
\end{table}

\end{document}